\documentstyle[preprint,eqsecnum,aps]{revtex}
\begin{document}
\newcommand{\cre}[2]{C^{\dag}_{#1\, #2}}
\newcommand{\ann}[2]{C_{#1\, #2}}
\newcommand{\num}[2]{n_{#1\, #2}}
\newcommand{\xh}{{\makebox{\boldmath$x$}}_{\rm h}}
\newcommand{\bfx}{{\makebox{\boldmath$x$}}}
\newcommand{\bfy}{{\makebox{\boldmath$y$}}}
\newcommand{\bfk}{{\makebox{\boldmath$k$}}}
\newcommand{\bfq}{{\makebox{\boldmath$q$}}}
\newcommand{\bfp}{{\makebox{\boldmath$p$}}}
\newcommand{\bfQ}{{\makebox{\boldmath$Q$}}}
\newcommand{\bfP}{{\makebox{\boldmath$P$}}}
\newcommand{\bfS}{{\makebox{\boldmath$S$}}}
\newcommand{\bfu}{{\makebox{\boldmath$u$}}}
\newcommand{\bfz}{{\makebox{\bf 0}}}
\newcommand{\Cos}{\cos \, }
\newcommand{\Exp}{\exp \, }
\newcommand{\eq}[1]{Eq.\ (\ref{#1})}
\newcommand{\eqs}[1]{Eqs.\ (\ref{#1})}
\newcommand{\fig}[1]{Fig.\ \ref{#1}}
\newcommand{\tabl}[1]{Table\ \ref{#1}}
\newcommand{\refer}[1]{Ref. \cite{#1}}
\draft

\title{Scaling properties of the ferromagnetic state \\
in the Hubbard model}

\author{K. Kusakabe$^*$ and H. Aoki}

\address{Department of Physics, University of Tokyo, Hongo, Tokyo 113, Japan}

\begin{abstract}
A numerical scaling analysis is used to show that Nagaoka's ferromagnetic state
in two-dimensional Hubbard model with one hole is supersede by an
antiferromagnetic (AF) state with a discontinuous jump in the total spin due to
the AF coupling as the Hubbard $U$ is made finite. The same applies to the
two-hole system, which has a spiral spin structure. We can show, via the
scaling, that the crossover to an AF state is a precursor of a pathological
coalescence of states having the minimum spin and Nagaoka's state at $U=\infty$
in the thermodynamic limit.
\end{abstract}


\narrowtext

To identify a stable itinerant ferromagnetism is one of the longest-standing
yet unresolved problems in the physics of strongly correlated electron systems.
Nagaoka's theorem is the first rigorous result, which served as a foundation of
subsequent studies.\cite{nagaoka,thouless,tasaki1}
Nagaoka's problem still holds a fascination as a remarkable manifestation of
the correlation effect in the doped Mott's insulator, where a key concept is
interference between the motion of a hole and its surrounding spin
configuration.\cite{trugman}
A recurring question on Nagaoka's ferromagnetism, however, is the singular
condition (the half-filled band doped with a single hole with the Hubbard
$U=\infty$): Is it possible to extend the ferromagnetism away from the extreme
condition down to finite number of holes and/or finite coupling strengths?

There have in fact been many attempts at looking into the possibility of
a finite ferromagnetic region in the vicinity of Nagaoka's limit
in the phase diagram.
Nagaoka has already shown that the spin-wave excitation in Nagaoka's
ferromagnet is singular in that the spin stiffness vanishes like the inverse of
the system size, $N$.
If one concentrates on the stability against one-spin flip (1SF) modes having
$S=S_{\rm max}-1$, however, the ferromagnetic phase seems to extend over a
finite area on the parameter space of the density of holes ($\delta =N_{\rm
h}/N$) and the antiferromagnetic (AF) coupling
($J=4t^2/U$),\cite{shastry,basile,edwards1,edwards2,hanisch}
but these variational (usually Gutzwiller-projected RPA) results
should be considered as an upper limit for the ferromagnetic boundary.
On the other hand, a recent high-temperature expansion result by Putikka {\it
et al.} shows that the true ground-state is significantly lower in energy than
the fully-polarized state for any hole density.\cite{putikka1,putikka2}
They also argue the possibility of a finite `ferrimagnetic' region, which is
identified from the uniform spin susceptibility diverging like the Curie law.

An exact way to explore the problem is the numerical diagonalization of finite
systems.
A serious point found in early studies is that Nagaoka's state is sensitive to
both the boundary condition and the number of holes,\cite{takahashi1,riera}
 where {\it singlet} states can supersede Nagaoka's state even at $U=\infty$.
These results have motivated subsequent
investigations.\cite{dw,fang,suto1,toth}
As for the instability of Nagaoka's state against $J$,
Ioffe and Larkin\cite{ioffe} have conjectured that a crossover to an AF phase
with a jump in $S$ can possibly occur for increasing $J$.
Nature of the lower-spin states that compete with Nagaoka's, however, is not
fully understood yet.

The purpose of the present paper is (i)to give the low-lying spectrum including
both charge and spin excitations, which will serve as a foundation (ii)to
obtain the scaling properties of the energy gap and level crossings to actually
identify how Nagaoka's state is superseded for the 2D square lattice.
In the formalism for (i), the heavy mass spin-wave, which causes the
coalescence of states comes from an interference between the charge (Bloch)
momentum and the spin momentum.
In (ii) we move on to calculate the low-lying states for every possible total
spin ($0$ or $1/2 < S < S_{\rm max}$) to
show that the first state to take over Nagaoka's as $U$ is decreased from
infinity is indeed an AF state rather than a partially polarized one, which
indicates a discontinuous jump in the total spin.
The same sudden crossover occurs for the $S=0$ ground state at $U=\infty$ that
appears for one hole with the anti-periodic boundary condition (APBC) or for
two holes with the periodic boundary condition (PBC), which we identify as a
{\it spiral} spin structure with a wave-length comparable to the system size.
The crossover by an AF state is related, via the scaling, to a pathological
degeneracy at $U=\infty$, where even the lowest-spin states become degenerate
with the Nagaoka's (or spiral) state in the thermodynamic limit ($N\rightarrow
\infty$).

In order to grasp the subtleness of Nagaoka's state, we start with introducing
the following formulation.\cite{ka1}
Consider the large-$U$ Hubbard model on a $d$-dimensional hyper-cubic lattice
with PBC.
An eigenfunction for $N_{\rm e}=N-1$ electrons with one hole can be written as
\begin{equation}
|\phi \rangle = \sum_{\xh} \Exp ({\rm i}\bfk \cdot \xh) \sum_{\{ \sigma_j \} }
f(\sigma_1 , \cdots , \sigma_{N_{\rm e}} ) \: \cre{\bfx_1}{\sigma_1}{} \cdots
\cre{\bfx_{N_{\rm e}}}{\sigma_{N_{\rm e}}}{} \rangle ,
\end{equation}
where $\xh$ is the position of a hole and
$\cre{\bfx_j}{\sigma_j}{}$ creates an electron with spin $\sigma_j$ at
$\bfx_j$.
The coefficient, $f(\sigma_1 , \cdots , \sigma_{N_{\rm e}} )$, can be regarded
as a spin wavefunction around the hole.
Note that we do not sum over the position of each electron:
A trick in writing the above formula is that the sequence ($\bfx_1,\ldots
,\bfx_{N_{\rm e}}$) (where no two positions coincide to exclude double
occupancies) of the multiplication of creation operators is fixed in such a way
that the creation at the same coordinate {\it relative to} $\xh$
enters in the same position for each value of $\xh$.
In this convention $f$ no longer contains $\xh$
reflecting the translational symmetry, while the factor $\Exp ({\rm i}\bfk
\cdot \xh)$
takes care of Bloch's theorem.

Consider the strong-coupling model, which is derived from the Hubbard model in
the second order in $1/U$-expansion.
Operating to the above basis we end up with a spin Hamiltonian,
\begin{eqnarray}
\cal{H}_{\rm spin} &=&
   - t \sum^d_{a=1} \{ \Exp (i\bfk \cdot \bfu_a ) \: {\cal{T}}_a
+ \Exp (-i\bfk \cdot \bfu_a ) \: {\cal{T}}^{-1}_a \}  \nonumber \\
& & + J \sum_{<i,j>} ({\bfS}_i \cdot {\bfS}_j - \frac{1}{4}) \nonumber \\
& & - \frac{J}{4} {\sum_{a,a'}}' \Exp (i\bfk \cdot \bfu_a + i\bfk \cdot
\bfu_{a'}) \:
{\cal{T}}_a {\cal{T}}_{a'}  \nonumber \\
& & \times ( \num{\bfu_a + \bfu_{a'}}{\uparrow}{} \num{\bfu_{a'}}{\downarrow}{}
+ \num{\bfu_a + \bfu_{a'}}{\downarrow}{} \num{\bfu_{a'}}{\uparrow}{}
- S^+_{\bfu_a + \bfu_{a'}} S^-_{\bfu_{a'}}
- S^-_{\bfu_a + \bfu_{a'}} S^+_{\bfu_{a'}}) \; ,
\label{spde}
\end{eqnarray}
which corresponds to the spin part of the $t-J$ model (including three-site
terms).
Here the operation, ${\cal{T}}_a (a=x,y,...)$, is roughly a translation of the
spin configuration along $a$-axis except for a spin on $\bfx_{\rm h}+\bfu_a$,
which is shifted onto $\bfx_{\rm h}-\bfu_a$ with $\bfu_a$ being a
nearest-neighbor vector.
The last line represents the three-site term, where the summation involves the
neighbor ($\bfu_a$) of the hole site and the third site ($\bfu_{a'}$) which is
the neighbor of $\bfu_a$.

The expression in the first term of \eq{spde} suggests the following picture.
Since ${\cal{T}}_a$ is almost a translation, it gives a phase which is roughly
the momentum of the spin configuration.
To be more precise, we can introduce,
in place of $f(\sigma_1 , \cdots , \sigma_{N_{\rm e}} )$ where spins are not
allowed to occupy the hole site ({\it i.e.} $\bfx_j \neq {\bfz}$),
another basis, $g(\bfy_1,\cdots ,\bfy_M)$, for $M$-spin-flip states on the
square lattice.
Here $\bfy_j$, which may be $\bfz$, denotes the position of $j$-th down spin.
We can then readily construct an eigenstate of the total momentum, $\bfQ$.
As a penalty for using the extended basis,
the operation ${\cal{T}}_a$ must be redefined: it gives zero if any $\bfy_j$
equals to $\bfz$ in $g$.
In this representation, the problem becomes a scattering of a spin at the hole
site.
Because the scattering is a higher-order process of $O(1/N)$ compared with the
translation,
we can obtain the asymptotic expansion in $1/N$
for the 1SF heavy-mass mode in 2D.
The solution exhibits a heavy mass (the band width $\sim 4\pi t/N\ln N$), which
coincides with the result by Barbieri {\it et al}.\cite{bry}

The advantage here is that the present formalism enables us to introduce a
quasi-momentum $\tilde{\bfQ}$ defined by
${\cal{T}}_a |\phi \rangle = \Exp (i \tilde{\bfQ}\cdot \bfu_a) |\phi \rangle$.
Then the asymptotic result indicates that $\tilde{\bfQ}$ is nearly equal to the
momentum, $\bfQ$, of the incoming state (with $|\tilde{\bfQ}-\bfQ |^2 \sim
O(1/N{\ln}N)$ for spin-wave excitations).
For a solution with $\tilde{\bfQ}({\bfk})$, the dispersion is given by
$\varepsilon({\bfk}) = -2t \sum_{a=1}^d \Cos (\tilde{Q}_a({\bfk})-k_a)$ for
each $\bfk$.
In a finite system we have discrete ${\bfk}$-points ($=\bfz,{\bfk}_1,\ldots)$
having single-particle excitation energy $\varepsilon(=0,\varepsilon_1,...$).
{}From this we can predict that the 1SF excitation from Nagaoka's state
should comprise a series of {\it bands},
each of which consists of an incoming wave, $\bfQ = \bfk + \bfk_i$ and has a
quasi-momentum $\tilde{\bfQ} = \bfk + \bfk_i + \Delta \bfk (i=0,...)$,
which are separated by the single-particle energy, $\varepsilon_i$,
with the band widths vanishing like $O(t/N{\rm ln}N)$.
Note that $\varepsilon_i$ itself vanishes like $O(t/N)$, which roughly gives
the gap
between heavy-mass bands.

This picture is in fact confirmed from the numerical diagonalization of a
290-site system in \fig{spinspect}.
Interestingly the above picture also holds
for two-spin-flip (2SF) excitations, where
the width of the lowest 2SF continuum as well as the width of the lowest 1SF
dispersion fit to
a size dependence of $\Delta_{1SF}(\infty), \Delta_{2SF}(\infty) \sim t/N{\rm
ln}N \ll \varepsilon_1 \sim O(t/N)$ within a finite-size correction (inset of
\fig{spinspect}).
The energy of 2SF mode is shown to be approximately twice that of 1SF, so that
the spin-waves are nearly free.
This is natural because the interaction between spin-waves is a higher-order
process than the scattering of a single spin at the hole site.

We now turn to finite $U$ for a fixed boundary condition in finite systems to
probe
level crossings between Nagaoka's state and other states having various total
spin.
So far Nagaoka's state is suggested to change to partially
polarized ones, then finally to a lowest-spin state after several level
crossings in 2D from numerical studies on 10- and 16-site
systems.\cite{kaxiras38,os1,barnes45,elrick}

However, if we increase the size to a 20-site system here, the strong-coupling
model ($t$-$J$ model with or without the three-site terms) shows the {\it
absence} of the intermediate partial polarization in the ground state,
(\fig{20crossing})
where the range of $J$ for the partially polarized region
decreases systematically to zero (\fig{jcscale}).
This suggests that an abrupt crossover from the fully spin-polarized state to
an unpolarized state occurs for larger systems.

We can relate this feature with the spectrum in the strong-coupling limit via
the scaling property of the level crossings.
We have previously reported from the value of $U_{\rm c}$ for the level
crossings that a 2SF mode takes over Nagaoka's state before a 1SF mode does
so.\cite{ka1}
If we plot here the crossing point, $U_{\rm c}^{\rm AF}$, between Nagaoka's
state and the AF state as well, we see that this is the crossing that comes
downwards most rapidly for increasing $J$ (inset of \fig{20crossing}).
Thus we obtain a lower bound for $U_{\rm c}^{\rm AF}$ as
$U_{\rm c}^{\rm AF} > \frac{4t}{\pi} N \ln N$.
The asymptotic form for the energy of the AF state is expected to be
\begin{equation}
E_{\rm AF}=-4t +\Delta_{\rm AF} - a N \frac{4t^2}{U} \; ,
\end{equation}
where $\Delta_{\rm AF}$ is the energy gap between the AF state above Nagaoka's
state at $U=\infty$ and the last term with $a \sim$ O(1) is the AF exchange
interaction.
Hence we end up with an upper bound for $\Delta_{\rm AF}$, via $E_{\rm
AF}(U_{\rm c}^{\rm AF})=E_{\rm Nagaoka}=-4t$, as
\begin{equation}
\Delta_{\rm AF} < a \pi t/\ln N \; ,
\end{equation}
which vanishes in the thermodynamic limit.
Therefor the huge degeneracy of the ground state in Nagaoka's limit is seen to
include even the AF state.

Although it is premature to discuss whether a spin-polaron picture or string
picture \cite{trugman,string} applies here,
we have confirmed that the lowest $S=1/2$ state has the energy that varies
approximately linearly with $J$ for $J_{\rm c}^{\rm AF} \leq J\alt 0.1t$ and
$S(\bfQ)$ that remains to be peaked at $(\pi, \pi)$, which implies that the AF
state experiences no level crossing prior to the crossing with Nagaoka's for
systems up to 26-sites. (See also \cite{bonca39}.)

As far as finite systems are concerned, we find that the {\it number} of holes
is crucial while the {\it concentration} of holes is ill-defined in the Hubbard
model at $U\sim \infty$.
As already noted by Riera and Young,\cite{riera} there are similarities among
the systems having the same number of holes.
Thus we can look for a scaling property in the series of finite samples with
the same number of holes.
For two holes, we can look at
the level crossing between the spiral-spin ground state for large $U$
characterized by $S(\bfQ )$ having four peaks around the $\Gamma$-point and the
AF state.
If we fit the crossover point from Nagaoka's state, $J_{\rm c}^{\rm AF}$
$(=4t^2/U_{\rm c}^{\rm AF})$ against $\delta$ with a power-low, we have
$J_{\rm c}^{\rm 1 hole} \propto \delta^{1.73},
J_{\rm c}^{\rm 2 hole} \propto \delta^{1.74}$.
Thus for $J \rightarrow 0$
Nagaoka's (or spiral) state is only realized for an infinitesimal doping, if
this scaling persists to $\delta \rightarrow 0$.
We notice in \fig{jcscale} that two scaling curves, although having similar
exponents, differ considerably, which endorses the difficulty in defining the
hole concentration.
Since both $J_{\rm c}^{\rm 1 hole}$ and $J_{\rm c}^{\rm 2 hole}$ are tangent to
the $\delta$ axis, a convex functional form for other number of holes would be
required to have a finite region for magnetism.
If we compare our scaling relation with the result (dashed line in
\fig{jcscale}) from the high-temperature expansion by Putikka {\it et al}, the
width for the partial polarization for one- and two-holes vanishes in what they
call is the ferrimagnetic region.
As far as finite systems with $N\sim 20$ are concerned, our result thus
indicates the absence of magnetism in this region.

However, since a finite number of holes only correspond to an infinitesimal
hole doping for $N\rightarrow \infty$,
the true problem of the ground state for finite concentrations of holes in the
thermodynamic limit remains an open question.

We wish to thank Prof. M. Ogata for illuminating discussions.  We are also
grateful for Prof. Y. Nagaoka in the early stage of the present work.

\begin{figure}
\caption{\label{spinspect}
Spin-wave excitations from Nagaoka's state (solid circle) in a 290-site 2D
square lattice with $U=\infty$.
Open circles represent one-spin-flip (1SF) excitations, while hatched regions
represent continuum of two-spin-flip excitations.
The arrow denotes the first single-particle excitation energy, $\Delta_{\rm
single}$.
The inset shows the size dependence of the band width, $\Delta_{\rm 1SF}$, of
the lowest 1SF mode (solid line), the width, $\Delta_{\rm 2SF}$, of the lowest
2SF continuum (broken line), and the bottom-to-bottom spacing of the first two
2SF continua, $W_{\rm 2SF}$ (chain line).
We also indicate the dependence, $4\pi t/N\ln N$ (lower dotted line),
and the single-particle energy $\Delta_{\rm single} (\sim 1/N$) (upper dotted
line).
}
\end{figure}

\begin{figure}
\caption{\label{20crossing}
Level crossings in a 20-site $t$-$J$ model in the small $J$ regime. The ground
state changes from Nagaoka's state to a singlet state at $J_{\rm c}^{\rm
AF}=0.0413t$.
The inset shows the scaling of various $U_{\rm c}$'s for $t$-$J$ models with
3-site terms defined in the text.
The dotted line represents the analytic evaluation,
$U_{\rm c}^{\rm 1SF}=(4t/\pi)N\ln N$,
while the dashed, chain and solid lines are a guide to the eye.
}
\end{figure}

\begin{figure}
\caption{\label{jcscale}
Scaling for the crossover point, $J_{\rm c}$, for the one-hole doped case
between Nagaoka's state and the antiferromagnetic (AF) state, and for the
two-hole doped case between the spiral-spin state and the AF one.
Bars represent {\it not} the error bar but the range over which partially
polarized states are realized.
A broken line is the boundary for the `ferrimagnetic' region obtained by
Putikka {\it et al.} (\protect{\refer{putikka1,putikka2}}).
}
\end{figure}

\end{document}